# Why honor heroes? The emergence of extreme altruistic behavior as a by-product of praisers' self-promotion

Jean-Louis Dessalles *

*LTCI, Télécom Paris, Institut Polytechnique de Paris, 19 Place Marguerite Perey, F-91120 Palaiseau, France*

ABSTRACT

Heroes are people who perform costly altruistic acts. Few people turn out to be heroes, but many spontaneously honor heroes by commenting, applauding, or enthusiastically celebrating their deeds. The existence of a praising audience leads individuals to compete to attract the crowd's admiration. The outcome is a winner-take-all situation in which only one or a few individuals engage in extreme altruistic behavior. The more difficult part is to explain the crowd's propensity to pay tribute from an individual fitness optimization perspective. The model proposed here shows how heroic behavior and its celebration by a large audience may emerge together. This situation is possible if admirers use public praise as a social signal to promote their own commitment to the values displayed by the hero.

The existence of individuals who risk their fitness and, at times, their lives to perform altruistic acts poses a challenge from a Darwinian perspective. Yet, this behavior becomes more comprehensible in the presence of an audience that rewards such acts with elevated social status and glory. In a world where langnage was not used to comment on the actions of others, there would be no glory. Individuals could at best gain the esteem of a few direct witnesses to their actions (Hawkes & Bliege Bird, 2002). Glory arises from the human tendency to comment on and praise the deeds of third parties. The present study aims to demonstrate, first, that extreme altruism can emerge in the presence of social praise and, second, that praise itself may function as a viable social signaling strategy. In summary, heroism and praise reinforce one another, serving as distinct yet complementary mechanisms for demonstrating social value. The following sections review a few studies on heroic behavior, with a focus on its relationship with praise, as well as research on praise and its potential role as a social signal. A social signaling model is then presented to explain the co-emergence of praising behavior and heroism, followed by its implementation in an agent-based simulation. The paper concludes with a discussion of the limitations and contributions of this research.

## 1. Why are there heroes?

High-risk or high-cost altruistic behavior is not uncommon among humans (Becker & Eagly, 2004; Kelly & Dunbar, 2001; Smirnov et al., 2007). It concerns all genders (Banks, 2019; Basak, 2015; Becker & Eagly, 2004; Bloom, 2011; Mama & Okazawa-Rey, 2012), notwithstanding authors' frequent tendency to give priority to male heroism by focusing on warfare (Kelly & Dunbar, 2001; Rusch et al., 2015; Smirnov et al., 2007). Heroic behavior is generally restricted to prosocial acts that expose the actor to the risk of dying or suffering serious physical consequences (Becker & Eagly, 2004; Goethals & Allison, 2012; Rusch, 2022; Stenstrom & Curtis, 2012). More inclusive definitions include social heroism, in which the costs are less dramatic but are certain to occur, often over extended time frames (Franco et al., 2011), as when nurses were called heroes at the beginning of the COVID-19 pandemic (Einboden, 2020). Risk-taking may even be optional, as when athletes are honored as "sports heroes" (Bifulco & Tirino, 2018).

As a case of extreme altruism among nonrelatives, heroic behavior is problematic from a Darwinian perspective: heroes seem to favor the propagation of others' genes at the expense of their own. Though some authors invoke group selection (Johnson, 1996; Lukaszewski et al., 2016) or kin selection (Smirnov et al., 2007), most studies link the existence of heroes to the presence of an audience ready to admire them (e.g. Goode, 1978; Hahl & Zuckerman, 2014; Kafashan et al., 2014, 2016; Rusch et al., 2015). The systematic appraisal of heroic acts contributes to the development of a hero's fame (Hahl & Zuckerman, 2014) and, indirectly, leads to status hierarchies where heroes often occupy dominant positions (Garfield et al., 2019; Gould, 2002). The prestige that emerges from public approval confers significant advantages and privileges on heroes (Henrich & Gil-White, 2000) and, as will be examined below, may be sufficient to explain why some individuals choose to perform risky altruistic actions. The other, more challenging side of the question is why people choose to bestow prestige, status, and honor upon heroes in the first place.

* Corresponding author.
*E-mail address:* dessalles@telecom-paris.fr.

## 2. Admirers' motives

Expressions of support for prominent figures are ubiquitous in human societies (Goethals & Allison, 2012; Klapp, 1949; Schrift, 2001). Their contemporary versions range from liking content on social media (Hess et al., 2022; Khan, 2017) to becoming activists or devoted fans (Chadborn et al., 2017; Thèberge, 2006). Public praise can be manifested formally during celebrations (De Regt, 2018), but also in a more mundane form during daily conversations (Algoe & Haidt, 2009). Conversational praise seems to be ubiquitous among human cultures, including in hunter-gatherer societies (Wiessner, 2014).

It could be argued that people are drawn to others' feats merely because such actions are, by definition, extraordinary. Human conversation is primarily oriented toward unexpected events (Dessalles, 2016) and admirable actions are just one instance of surprising occurrence among many. However, comments on heroic feats are rarely limited to expressing surprise. On the one hand, most gossip about exploits is accompanied by appraisals, with a significant proportion of these being positive (Ben-Ze'ev & Goodman, 1994; Levin & Arluke, 1985; McDonald et al., 2007). On the other hand, heroic acts are not mere unexpected accomplishments like those found in the *Guinness Book of Records*. They are typically characterized by intentional risk and altruism, performed to earn public approval.

As noted by Kafashan et al. (Kafashan et al., 2016), overt judgments of heroism may involve marginal costs that, although small in comparison to the sacrifices made by heroes, are not always negligible. Public displays of support, such as participating in celebrations or defending heroes in discussions, require time and energy. In addition, granting elevated status and accompanying privileges to heroes can result in lesser status and reduced access to valuable resources for the judge. It may also put one's social capital at risk (e.g. supporting NSA whistleblower Edward Snowden may limit access to job opportunities). Mere positive gossip, by revealing one's opinions, may sometimes have negative consequences (but also positive ones, as suggested in the present study).

How does expressing admiration bring a return on investment? Marks of deference might be seen as a way for admirers to approach competent models and get some support in return (Gould, 2002) or learn from them (Henrich & Gil-White, 2000). However, ethnographic data do not provide strong support for the social learning component of the prestige model (Garfield et al., 2019). Moreover, admiration is often expressed toward unattainable targets that may not be aware of specific supporting acts (Bifulco & Tirino, 2018). Granting prestige and esteem could be a way of giving heroes control over collective decisions (Chapais, 2015; Garfield et al., 2019), e.g. by letting them punish uncooperative acts (Redhead et al., 2021). Esteem can also be withdrawn and heroes denigrated, as a way for supporters to keep control over them (Goode, 1978; Hahl & Zuckerman, 2014). Investing in praise to control who controls is unlikely to be worth the cost, though. As for any invocation of collective benefit, the return on marginal investment is divided by the number of supporters. Why spend time and energy singing the praises of some hero if the rest of the audience does it for us? Alternatively, the readiness to praise third parties could be understood as part of a reciprocal exchange of valuable information about capable individuals. However, reciprocal exchanges of valuable information are exposed to the problem of free-riding; it must rely on precise bookkeeping of the informational value of each contribution, which is not confirmed by observation (Dessalles, 2020). Furthermore, reducing praise to its mere informational content does not account for the emphasis and enthusiasm often displayed when it is expressed. The present study offers a novel perspective, in which praise functions as a social signal.

## 3. Praise as a second-order signal

After Greta Thunberg rose to prominence in 2018, politicians sought to be photographed with her to demonstrate their ecological commitment (Murphy, 2021). Similarly, speaking positively about her during conversations became a means of signaling concern for climate change. For some authors, speaking well of prominent figures is a way of drawing their attention to oneself (Anderson et al., 2020; Gould, 2002). The central hypothesis of the present study is that the prime target of praise is not the "hero" but rather the immediate witnesses of the praising act (Fig. 1). As Tannen puts it: "Talking about someone who is not there is a way of establishing rapport with someone who *is* there." (Tannen, 1990, p. 107). Observations show that positive gossip correlates with friendship ties among colleagnes (Grosser, Lopez-Kidwell, Labianca, & Devillard, 2010) and children (Lansford et al., 2006). The primary motivation for displaying fan identity (Chadborn et al., 2017) or participating in patriotic celebrations (De Regt, 2018) is not entertainment, learning, or self-enhancement, but rather the pursuit of new friendships.

How is praise conducive to making and keeping friends? According to Tannen: "By agreeing about their evaluation of someone else, people reinforce their shared values and world views." Individuals tend to affiliate with people who share their values. Social network algorithms favor this tendency by leveraging common 'likes' to make predictive recommendations about friendships (Agarwal & Bharadwaj, 2013). Values become apparent only when one performs actions in line with them. For example, a climate-conscious individual might choose to carpool and offset the carbon footprint of a trip. However, such actions are visible only when they are performed. This is where praise comes into play. Similar to Facebook's 'thumb-ups', praise serves as a way to signal one's values. For instance, supporting Greta Thunberg's recent actions provides an opportunity to discuss one's own contributions to climate efforts while remaining conversationally relevant (Dessalles, 2016).

From this perspective, praising is a social signal, i.e., a signal used to make friends. Since it is about feats which are social signals themselves (Kafashan et al., 2016), one may call it a *second-order signal*, a signal about a signal. Blame, as the mirror image of praise (Jordan et al., 2016; Lie-Panis & Dessalles, 2023), is also a second-order signal. At the collective level, blame leads to social signal leveling (Lie-Panis & Dessalles, 2023). The following model demonstrates how positive second-order signals, in contrast, can lead to the opposite effect, where a few individuals distingnish themselves by displaying extreme signals.

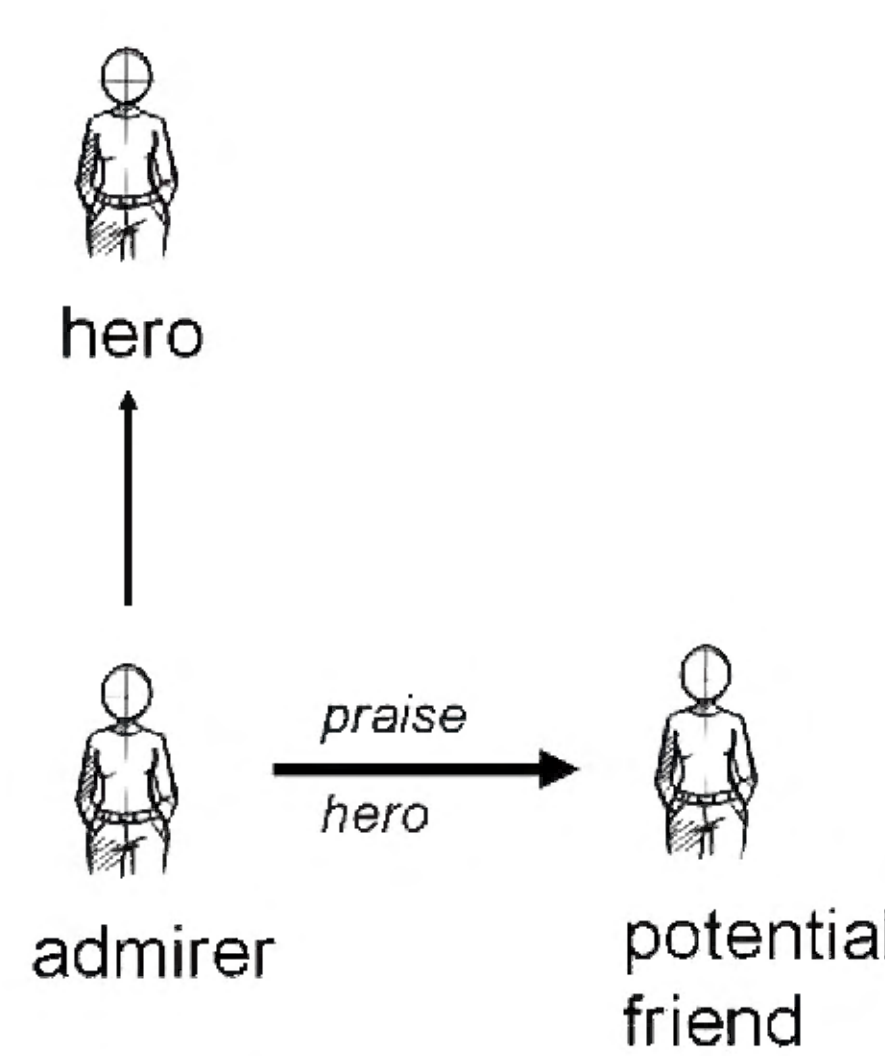

**Fig. 1.** The prime target of praise is not the "hero".

## 4. Model

The model described here shows that a few individuals can benefit from adopting high-cost behavior within a population where praise is used to advertise one's social value. Consider a population of $N$ agents that differ by one hidden characteristic $q$ (e.g. altruism or bravery). They may advertise it by sending a costly signal $S(q)$ (e.g. performing altruistic acts or risky feats). For the sake of simplicity, characteristics $q$ are supposed to be uniformly distributed over the range [0,1]. Signal cost $C(S,q)$ is supposed to increase with $S$ (its partial derivative $C_1$ is positive) and to decrease with $q$ ($C_2 < 0$). With the latter assumption, competence $q$ appears as the readiness/ability to emit the costly signal. Let's call it *competence*. In addition to the signaling strategy $S(q)$, agents can choose whether to praise a top signaler or remain silent.

Agents meet systematically in pairs for two rounds. During the observation round, agents see each other's signal with probability $p_1 \leq 1$. This limited visibility reflects the fact that the opportunities to demonstrate the signal (e.g. through actions) are few and far between, with only a few witnesses. The restriction to dyadic encounters is intended to prevent individuals from observing everyone at once, though it could be partially relaxed. During the communication round (Fig. 1), agents can choose to praise the best signaler they encountered during the first round (referred to as their "hero"), incurring a cost $c_p$. Praising makes the hero's signal visible to the potential friend (no deception). Additionally, the potential friend will also observe the admirer's signal with probability $p_p$. This reflects the idea that the admirer has an opportunity to draw attention to their own actions or, at the very least, their own values. The potential friend then decides whether to affiliate with the praiser based on the praiser's signal (if visible). An agent may affiliate with up to $m$ signalers and attract up to $k$ followers. These limitations represent the bounded amount of time individuals can spend with each other (Dessalles, 2014).

After these two rounds of encounters, payoffs are computed. Agents receive a payoff $F$ for each affiliate they attract, with the maximal social payoff being $k \times F$. Let us first examine the basic social game (without praise), before analyzing the effect of praise on the emergence of "heroes".

### 4.1. Basic social signaling (without praise)

In a situation of perfect visibility ($p_1 = 1$), the best signalers are certain to attract $k$ followers each. The system evolves toward a situation in which the most competent individuals $q > \widehat{q}_1$ emit the same signal $S_1$, while other individuals don't invest in the signal whatsoever (Fig. 2).

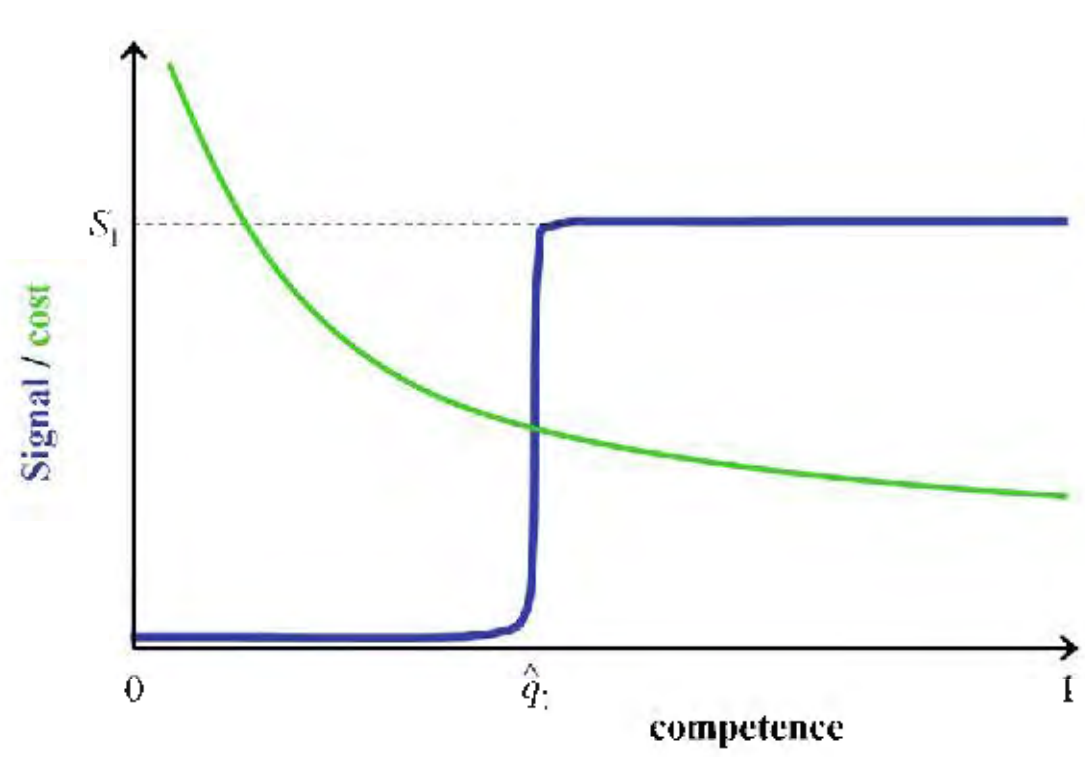

**Fig. 2.** Binary signaling at equilibrium. The blue curve shows the signal level $S(q)$ (competence $q$ is supposed to be uniformly distributed between 0 and 1). Signal cost $C(S,q)$ for a given signal intensity $S$ is proportional to the green curve (signals are cheaper as $q$ gets closer to 1).

This all-or-nothing signaling pattern is an evolutionarily stable strategy (ESS) (see supplementary material). It is a situation of honest social signaling (Gintis et al., 2001), here generalized to continuous values. If $m < k$, then $\widehat{q}_1 = 1 - \frac{m}{k}$.

If $p_1 < 1$, the situation of Fig. 2 still holds, though with a different value of $\widehat{q}_1$. If the expected number $p_1 \times N$ of individuals who witness a given signal is such that $m < p_1 N < k$, then $\widehat{q}_1 = 1 - \frac{m}{p_1 N}$. If we note $f(p_1) = min(p_1 N, k)$ the average number of followers per signaler, the signalers' benefit is: $f(p_1) \times F - C(S_1, q)$. The signal $S_1$ displayed by more competent individuals is such that there is no incentive for a less competent individual to compete. $S_1$ stabilizes at a level that makes the benefit equal to zero at the threshold (see supplementary material). It is the solution of:

$$C(S_1, \widehat{q}_1) = f(p_1)F. \tag{1}$$

### 4.2. Signaling in the presence of praise

Praise may seem like a peculiar behavior. While it enhances the reputation of the admired individual, it also invites comparison with the praiser's often inferior qualities. However, the model demonstrates that praising behavior can still emerge, not as a means to please the explicit target, but rather as a strategy to attract potential friends who are more accessible. Agents may increase the visibility of their signal from $p_1$ to $p_2 \approx p_1 + p_p$ by selecting a top-signaler to praise. Praising acts generate a side-effect, which translates into a bonus $s$ for the praised individual. This bonus represents the contribution of praise to the hero's fame. It is important to note that praises are not social links in the model. Praising (often inaccessible) heroes is by essence an asymmetrical relation, allowing a "hero" to receive an unbounded number of praises. It is distinct from real affiliations among friends which require spending time together. At equilibrium, non-signalers have no incentive to praise, as this would only make their absence of signal more visible. Signalers will systematically praise as soon as the cost of praising $c_p$ satisfies:

$$(f(p_2) - f(p_1)) \times F > c_p. \tag{2}$$

When praise makes rarely observed events ($p_1$ small) visible ($p_2$ significant), this condition is simply: $(k - p_1 N)F > c_p$. The existence of praise moves the signaling threshold from $\widehat{q}_1$ up to $\widehat{q}_2 = min\left(1 - \frac{m}{p_2 N}, 1 - \frac{m}{k}\right)$ and the signal from $S_1$ to a new value $S_2$ obtained by substituting $p_1$ for $p_2$ in (1):

$$C(S_2, \widehat{q}_2) + c_p = f(p_2)F. \tag{3}$$

The total windfall to be shared among "heroes" is $P = N(1 - \widehat{q}_2)s$. Competition for attracting praise may lead to a winner-take-all situation in which the most competent individual sends an intense signal $S_H$ to win $P$ (Roberts, 1998).

A situation in which $H$ mostly competent individuals ("heroes") send the same signal $S_H$ and receive the same amount of praise $\frac{P}{H}$ is however possible (Fig. 3). This multi-hero situation ($H > 1$) occurs if increasing the signal level by a minimal noticeable relative value $\delta$ involves a prohibitive cost for the top-most individual:

$$\left(P - \frac{P}{H}\right) < C(S_H \times (1 + \delta), 1) - C(S_H, 1). \tag{4}$$

If $\delta$ is small enough, one gets a first-order approximation for this condition:

$$P\left(1 - \frac{1}{H}\right) < \delta S_H \times C_1(S_H, 1). \tag{5}$$

One may suppose that the signal cost increases asymptotically as the signal approaches a limit $L$ (e.g. available time, acceptable risk) and that performers measure the relative distance to that limit when evaluating costs. With these assumptions, the cost first derivative close to the limit $L$ has the form $C_1(S,1) = \frac{c_0}{L-S}$, where $c_0$ is a coefficient that controls the steepness when approaching $L$. A condition for observing $H > 1$ "heroes" is (see supplementary information for details):

$$S_H \geq \frac{L}{1+\frac{c_0\delta}{P\left(1-\frac{1}{H}\right)}}. \tag{6}$$

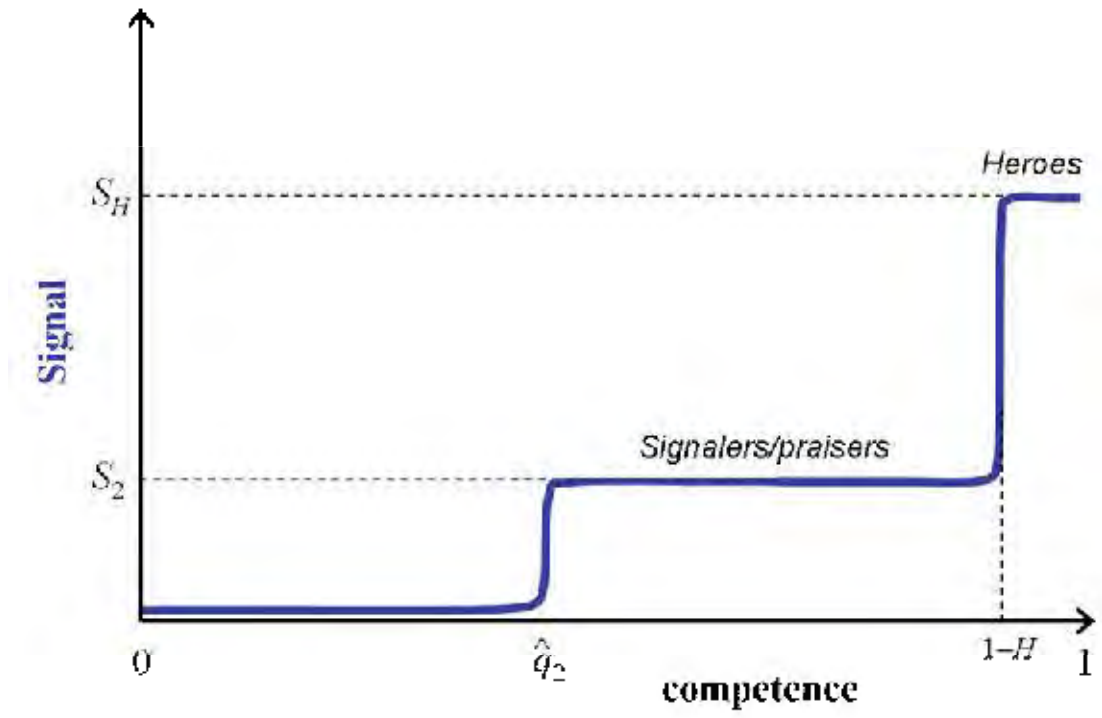

**Fig. 3.** Emergence of three signal levels depending on competence.

The number of "heroes" is obtained by observing that the least competent/motivated among them should have a positive benefit:

$$kF+\frac{P}{H} \geq C\left(S_H, 1-\frac{H-1}{N}\right). \tag{7}$$

If (7) is not satisfied for $H = 2$, there can be only one "hero" whose signal $S_{H-1}$ is just costly enough to discourage the second-most competent individual:

$$C\left(S_{H-1}, 1-\frac{1}{N}\right) \gtrsim kF+\frac{P}{2}. \tag{8}$$

Otherwise, one can plug the lower bound of (6) into (7) to get a constraint on $c_0$ to get more than one "hero" (see supplementary information):

$$kF+\frac{P}{H} \geq \frac{c_0}{1-\frac{H-1}{N}} log\left(1+\frac{P\left(1-\frac{1}{H}\right)}{c_0\delta}\right). \tag{9}$$

Fig. 4 shows that based on constraint (9), the number of "heroes" grows with the population's size for a given cost coefficient $c_0$. The simulations detailed below confirm that praising and heroism co-evolve until signals stabilize around three distinct levels: 0, $S_2$, and $S_H$.

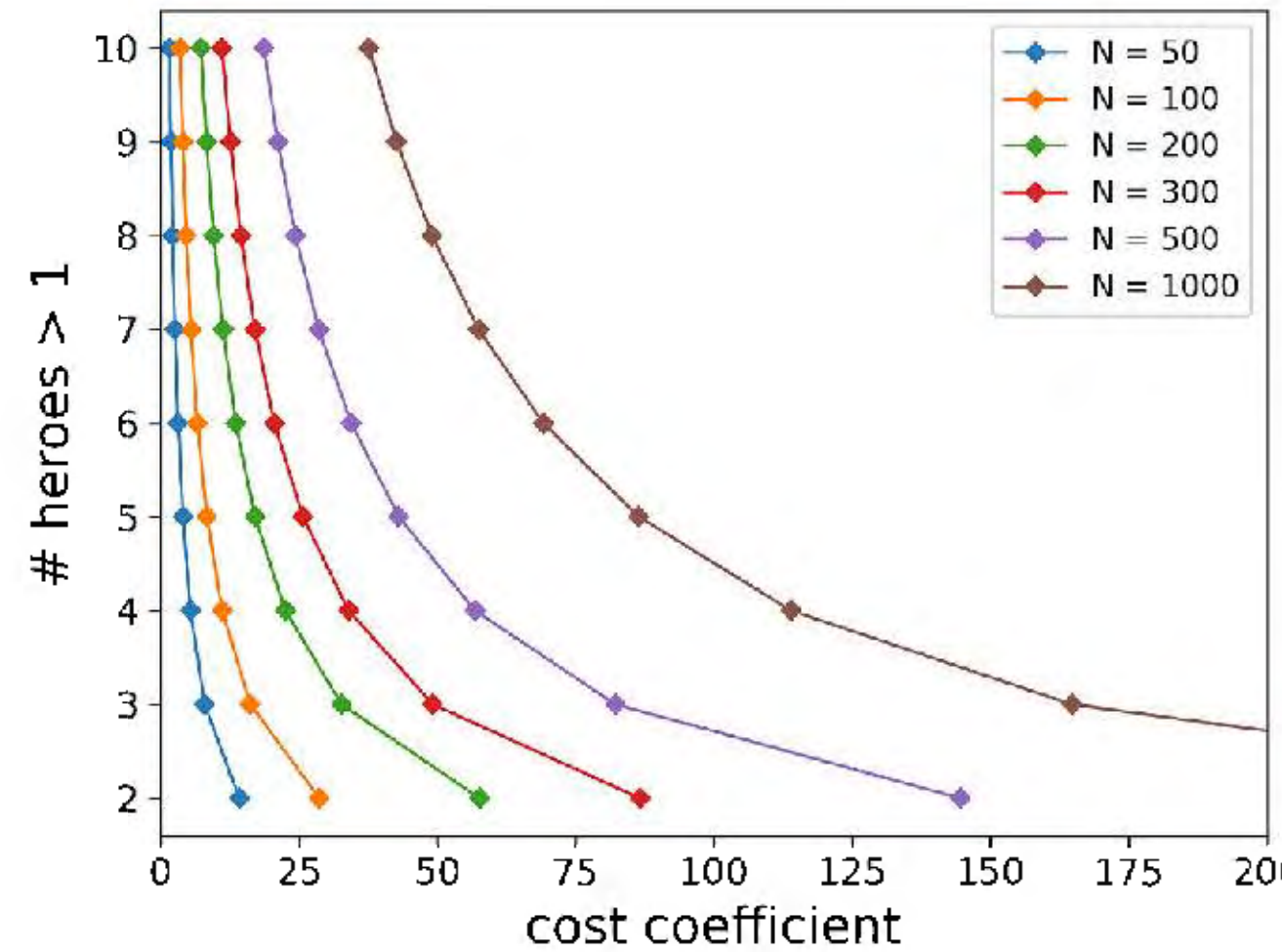

**Fig. 4.** Number of "heroes" when more than one can coexist. Signaling cost is proportional to the cost coefficient $c_0$ (x-axis). Marginal cost is supposed to grow asymptotically when the signal gets close to a limit $L$: $C_1(S,q) = \frac{c_0}{q(L-S)}$. One can see that larger populations (here $50 \leqslant N \leqslant 1000$) can support more "heroes". Parameters are $k = 5$, $F = 10$, $p_2 = 0.3$, $s = 3$ (hence $P = 1.2N$), $\delta = 0.1$, and $L = 100$. When $c_0$ is too large to fall within the curve (for a given $N$), there can be only one "hero".

## 5. Simulation

The purpose of implementing the model in an agent-based simulation is to assess its robustness in situations where matching among friends is imperfect. In both the simulation and real life, individuals may abandon some friendships in favor of more appealing alternatives. These former friends may take time to establish new affiliations. As a result, social matching remains incomplete, with good signalers ending up with fewer than $m$ accepted affiliations or less than $k$ affiliates.

Agents' behavior is controlled by two flexible behavioral traits that agents learn: the intensity of their signal and their probability of praising some "hero". These traits are initially set at 0. Agents may try out other values of the traits, mostly locally and, with a small probability, by making a significant change. They keep a limited memory of the most profitable values and restart their exploration from them. An agent dies when reaching maximum age and is reborn as a blank agent with the same competence. As in the model, interactions are kept local: individuals observe each other's signals during dyadic encounters with probability $p_1$ and follow the best signaler available. The structure of interactions and the calculation of payoffs follow the model's specifications.

Fig. 5 (a) shows the signal intensity $S_H$ attained by top-signalers. It reaches values close to the maximal asymptotic limit $L = 100$, either within large groups or when the payoff $s$ for being praised is significant. One can see in Fig. 5 (b) that individuals are more prone to invest in praising third parties when the default visibility $p_1$ of their own signal is smaller than 0.2. Fig. 6 confirms that the average number of "heroes" grows with group size.

## 6. Discussion

This study offers a proof of concept to show that heroism may emerge from the existence of a praising audience, while praising itself appears as a second-order social signal. Individuals praise heroes to draw attention to their own signal, so as not to be mistaken for non-signalers. With these simple assumptions, the population splits into three discrete categories: non-signalers, signalers, and super-signalers ("heroes") who are praised by signalers. The crucial assumption underlying the phenomenon is that some social signals are not fully visible or accessible, and that praising increases their visibility.

The model does not claim to capture the complexity of real situations. The competence underlying any real-life performance cannot be reduced to a mere number, as it depends on complex situational characteristics (strength, skill, training, experience, boldness) and on the motivation to prioritize signaling over addressing more immediate needs. Also, the model considers one single social dimension valued by all. Human groups establish social bonds based on a variety of values, such as sports, musical performance, and even "trendiness". Though the signal sent by "heroes" amplifies the signal sent by "normal" people along some dimension, the model could be extended to deal with several values simultaneously.

Signals and praise are considered truthful in the model because of their cost (Zahavi, 1975). A more realistic representation would include hypocrisy as a negative signal (as in Lie-Panis & Dessalles, 2023). The model is also agnostic about psychological motivations. Admiration may be accompanied by intense subjective experience (respect, joy, inspiration, awe, sympathy, enthusiasm, humility…). The model only

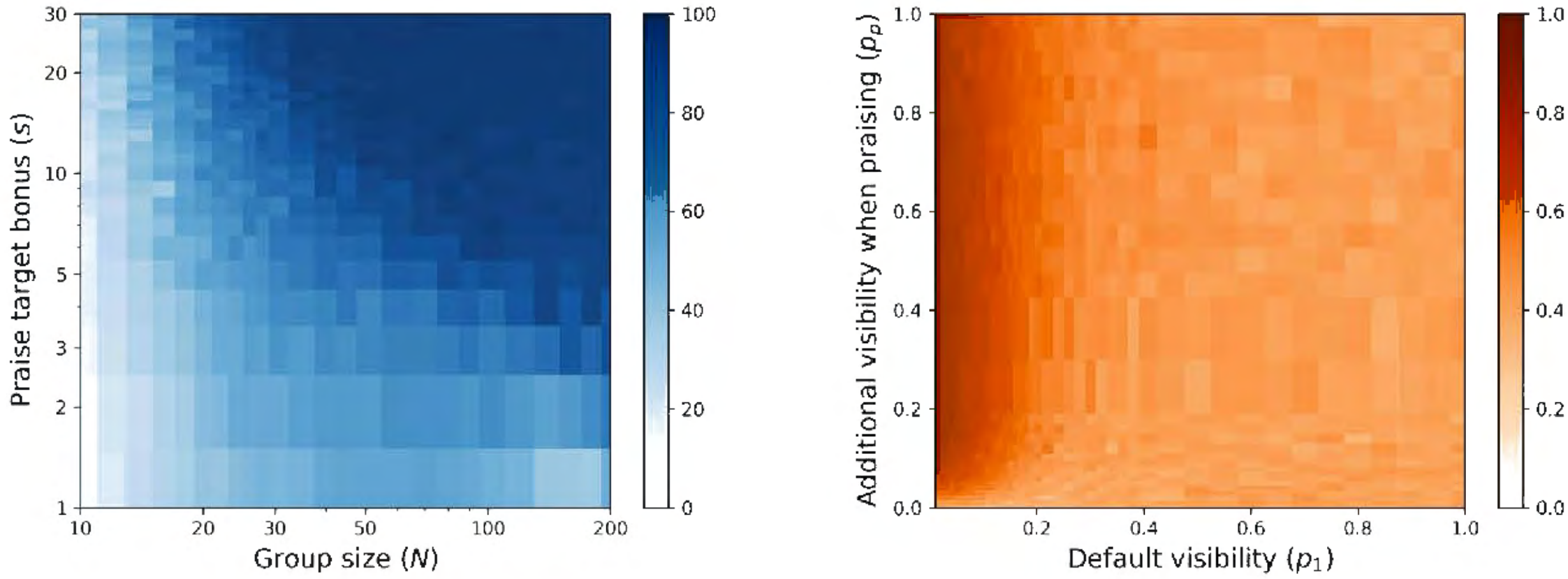

**(a) top signal intensity** **(b) praising probability**

**Fig. 5.** Simulation results. (a) Intensity of signals emitted by the top signalers, depending on group size $N$ and the bonus $s > 0$ received by praised individuals. (b) Average praising probability among signalers, depending on baseline visibility ($p_1$) and the additional visibility ($p_p$) due to the praising act. Default parameters are $k$ 5, $m$ 2, $F$ 10, $c_0$ 100, $c_p$ 3, $L$ 100; In (a): $p_1$ 0.1, $p_p$ 0.3; In (b) $N$ 50, $s$ 3.

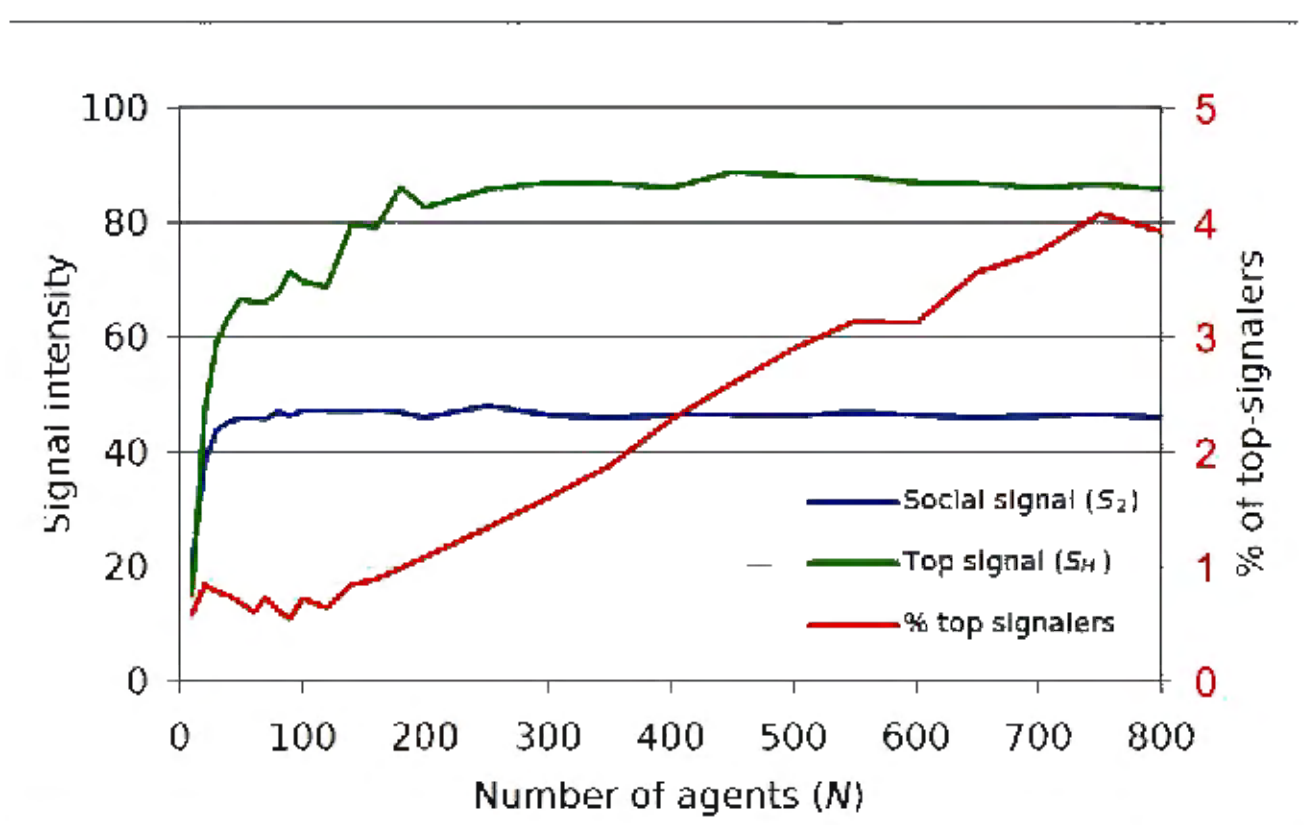

**Fig. 6.** Emergence of several "heroes" in large populations. Number of top signalers ("heroes") (in red, right scale), shown with the signal sent by these "heroes" ($S_H$) (green) and "non-heroes" ($S_2$) (blue) as a function of population size ($N$). Under the conditions of this simulation, one expects three "heroes" on average in a group of 500. Parameters are the same as for Fig. 5, with $c_0$ 100. (For interpretation of the references to colour in this figure legend, the reader is referred to the web version of this article.)

addresses the public aspect of admiration. However, the reason why praising feats may be associated with intense emotions becomes less mysterious if admiration has a social signaling function.

The paper proposes a mechanism that solves two mysteries at once: the disproportionate investment by certain individuals in altruistic behavior, and the existence of an audience willing to expend time and effort to commend their actions. Both behaviors can be understood as forms of social signaling: extreme altruists signal their virtues to elicit praise, while those offering praise signal their altruistic orientation to attract social bonds. This dual-signaling framework is stable and resistant to free-rider issues. It also explains how humble achievers (Hawkes & Bliege Bird, 2002) or unwilling heroes can still attain status. In the latter case, individuals who never claimed to have performed heroic acts ('All I did was my duty') may nevertheless be acclaimed by the crowd (Garcia, 2003, p. 238). In the model's logic, their actions are elevated to heroic status when praising individuals use them as an opportunity to display their own patriotism.

The significance of second-order social signals should not be underestimated. While negative evaluations lead to signal leveling (Lie-Panis & Dessalles, 2023), positive evaluations offer an opportunity to a few individuals who issue exaggerated signals to garner widespread approval. Such individuals, particularly heroes, are deeply invested in their reputations, with some even risking their lives to defend their 'honor.' In more everyday contexts, second-order signals help to better understand phenomena such as the rise of influencers on social media or the role of positive gossip in conversations. Sociological studies have often treated praise at face value, assuming it is what it appears to be. As suggested by the present study, though, fervent cheers at a concert broadcast live in a fan zone are not intended for the distant performer on stage, no more than a eulogy is for the deceased. Positive evaluations are never socially neutral; rather, they serve as strategic displays of the values we wish to share with others in our social circles. This new understanding of the function of praise raises interesting research questions. One is to understand why heroes may, in some cases, go as far as choosing to sacrifice their lives for the group (Whitehouse, 2018). Another is how the model might apply to contexts where multiple, potentially contradictory competencies are simultaneously valued in a population, possibly leading to societal segmentation.

## CRediT authorship contribution statement

**Jean-Louis Dessalles:** Writing – review & editing, Writing – original draft, Software, Methodology, Formal analysis, Conceptualization.

## Acknowledgements

The authors would like to thank Julien Lie-Panis for initial discussions about the model and Pat Barclay for his insightful comments on an earlier draft. This work was supported by Telecom-Paris.

## Appendix A. Supplementary data

Supplementary data to this article can be found online at https://doi.org/10.1016/j.evolhumbehav.2025.106656.

Supplementary Information for

# Why Honor Heroes? Praise as a Social Signal

Jean-Louis Dessalles*[1]

[1] *LTCI, Télécom Paris, Institut Polytechnique de Paris, 91120 Palaiseau, France*

May 24, 2024

## Contents

## S-1 First-order signals

We consider a group of $N$ agents differing by a hidden characteristic $q$ called *competence*. For the sake of simplicity, we suppose that competence is uniformly distributed between 0 and 1. Agents' competence controls their ability/motivation to invest in a costly signal $S(q)$; this means that sending $S(q)$ comes with a cost $C(S, q)$ where the partial derivative in $S$, $C_1(S, q)$, is positive while the partial derivative in $q$, $C_2(S, q)$, is negative.

Individuals affiliate to each other based on these signals. An agent may affiliate to $m$ other individuals and receive up to $k$ affiliations (these limitation results for instance from time constraints). When choosing whom to follow, an individual selects the best signaling agent among the available ones. Note that social competition only exists if $m < k$. Agents get payoff $F$ for each follower they could attract.

One original hypothesis of the model is that signals are directly visible only with probability $p$ (e.g. they correspond to punctual feats with few witnesses). Unobserved signals are considered to have zero value by default. If the signals of all encountered individuals happen to remain unobserved, selection for affiliation is performed at random.

### S-1.1 Emergence of a signaling threshold

Individuals publicly differ only by the signal $S(q)$ they send. We show (1) that at equilibrium, $S(q)$ is an increasing (non-decreasing) function of $q$; (2) that $S(q)$ is more precisely a step function that is zero below a threshold $\hat{q}$ and constant above; and (3) that this situation is evolutionarily stable.

(1) The optimal signal $S(q)$ sent by an individual with competence $q$ is an increasing function: $S'(q) \geq 0$. Consider two agents with competence $q_1$ and $q_2$ such that $q_1 < q_2$. If $S(q_1)$ is optimal for the first agent, a lesser signal $S(q_2) < S(q_1)$ cannot be optimal for the more competent agent. The latter would be better off sending $S(q_1)$ as well, since individuals' social payoff depends only on their signal and the signaling cost decreases with competence: $C(S(q_1), q_2) < C(S(q_1), q_1)$.

*Email: dessalles@telecom-paris.fr; ORCID: 0000-0002-3910-4611

(2) Could it be that all individuals send the same signal $S$? If $C(S, 0) < F$, all agents, including the least competent one, would attract one affiliate and get a positive payoff. This situation is however unstable as soon as there is social competition (see S-1.3). If (as we suppose) $C$ is a continuous function, the most competent individuals can raise their signal to a new value $S'$ such that $C(S', 0) > F > C(S', 1)$. They will attract new affiliates that the least competent individuals, who are now better off sending no signal, no longer get. A competence threshold $\hat{q}$ emerges from the signaling competition below which individuals don't send any signal.

With the same reasoning, we can see that the most competent individuals will increase their signal $S$ as long as they can take followers from the less competent. Suppose for now that top-signalers fill up with followers and attract $k$ of them. Any individual with competence $q$ such that $C(S, q) < k \times F$ will send the same signal $S$ to get $k$ followers as well. It would be sub-optimal for them to send a lesser signal $S' < S$ if it would mean losing $F$, and if they still attract $k$ followers with $S'$ then it would be sub-optimal for top-signalers to send $S$ instead of $S'$. We conclude that signals sent in the population evolve towards a step function (Figure S1). The $m \times N/k$ most competent individuals will attract all $m \times N$ available affiliations (for the sake of simplicity, we suppose that $mN$ is divisible by $k$). Since competence is supposed to be uniformly distributed, we deduce that $\hat{q} = 1 - m/k$.

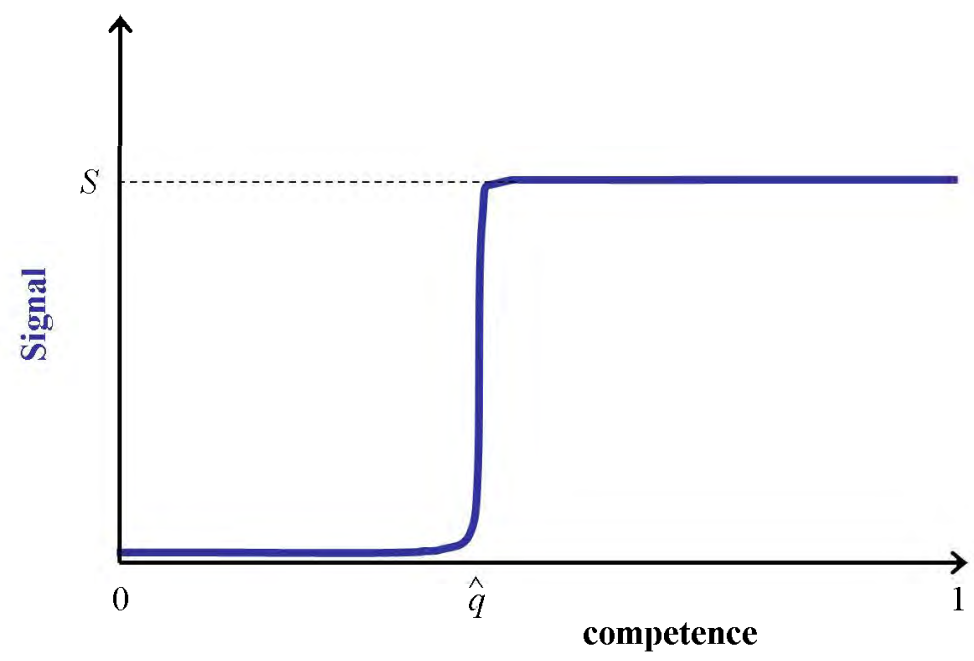

Fig. S1: Signal level $S(q)$ evolves towards a step function.

To determine $S$, we observe that its cost is affordable for an individual at $\hat{q}$, but not for an individual at $\hat{q}_- = \hat{q} - 1/N$ just below in the order of competence:

$$C(S, \hat{q}_-) > kF \geq C(S, \hat{q}). \tag{S-1.1}$$

If $N$ is large enough, we can make a first-order approximation based on the width of the interval:

$$C(S, \hat{q}_-) - C(S, \hat{q}) \approx -C_2(S, \hat{q})/N \gtrsim 0. \tag{S-1.2}$$

Equation (S-1.2) shows that signaling cost $C(S, \hat{q})$ cancels out social payoff $kF$ for individuals at $\hat{q}$ at soon as $N$ is significant. The value of $S$ can be deduced if we write that benefit is close to zero at $\hat{q}$:

$$C(S, 1 - \frac{m}{k}) \approx kF. \tag{S-1.3}$$

(3) We just showed that the situation depicted in Figure S1 corresponds to a Nash equilibrium. It is trivially an evolutionarily stable strategy (ESS) since the situation will go back to its stable state, no matter how many agents initially deviate from it.

## S-1.2 Low signal visibility

Two phenomena may limit the number of affiliates a signaler may attract. One of them is the low visibility of signals. If $p$ is small enough, fewer than $k$ individuals might see a signaler's performance. The reasoning leading to (S-1.3) remains valid. We just replace $k$ by the expected number $p \times N$ of affiliates:

$$C(S, 1 - \frac{m}{pN}) \approx pNF. \tag{S-1.4}$$

This situation presupposes that $p > m/N$. What happens if $p < m/N$? Visibility $p$ is too low for signalers to attract all $m \times N$ available votes. There seems to be no incentive to signal. A situation where everyone sends the same signal $S$ (which could potentially be $S = 0$) and receives $m$ random votes would not, however, constitute a Nash equilibrium. Any individual could attract $pN + (N - 1 - pN)m/N \approx pN + m$ votes on average just by sending a signal $S' > S$. Signals will rise until a signaling threshold $\hat{q}$ emerges. Signalers get $pN$ votes on average in addition to the $m - pN(1 - \hat{q})$ random votes. Signaling cost balances the payoff difference $pNF$ between signaling and not signaling at the threshold:

$$C(S, \hat{q}) = pNF. \tag{S-1.5}$$

(S-1.5) is no longer sufficient to determine both $S$ and $\hat{q}$, but the signaling threshold $\hat{q}$ is uniquely determined once $S$ is fixed. We can deduce a function $\hat{q} = g(S)$ from (S-1.5). The value of the signal $S$ sent by signalers is controlled by top-signalers, who optimize their benefit:

$$C_1(S, 1) = \frac{d}{dS}[pN + m - pN(1 - g(S)]F = pNg'(S)F. \tag{S-1.6}$$

Using implicit function theorem, we get:

$$C_1(S, 1) = -pNF\frac{C_1(S, g(S))}{C_2(S, g(S))}, \tag{S-1.7}$$

$$\frac{C_1(S, 1)}{C_1(S, g(S))} = \frac{pNF}{-C_2(S, g(S))}. \tag{S-1.8}$$

With additional hypotheses concerning the cost function (see section S-2.2), we might rewrite (S-1.8) as:

$$g(S) = \frac{pNFg(S)^2}{C(S, 1)}, \tag{S-1.9}$$

which gives:

$$g(S) = \frac{C(S, 1)}{pNF}. \tag{S-1.10}$$

$g(S)$ increases from 0 to 1. We will suppose (section S-2.2) that $C(S, 1)$ increases from 0 to $\infty$, with a linear start. If $N$ is large enough, (S-1.10) has a unique non-zero solution.

### S-1.3 Encounters and affiliations

Another phenomenon might affect social competition. In the preceding section, we implicitly supposed that social matching is perfect: signalers are guaranteed to recruit followers among individuals who can see their signal. This requires that individuals have a complete memory of all signalers they have previously encountered and are free to join them afterward. We might want to relax this hypothesis, as we did in our implementation. When individuals observe signals and select signalers in the same move, social matching might end up imperfect. Even when all individuals meet systematically, signalers may be seen by individuals who are already following $m$ other signalers. They find themselves alone, despite having sent a proper signal. As shown in Figure S2 (a), the number $f(p, \hat{q})$ of followers a signaler might expect when social matching is imperfect is smaller than the theoretical maximum $f(p) = min(pN, k)$.

Supposing $pN < k$, we must rewrite (S-1.4) as:

$$C(S, \hat{q}) = f(p, \hat{q})F. \tag{S-1.11}$$

As previously, we deduce from (S-1.11) a function $g$ such that $\hat{q} = g(S)$. Top-signalers optimize their benefit:

$$\frac{\partial}{\partial S}f(p, g(S))F = C_1(S, 1). \tag{S-1.12}$$

We can rewrite (S-1.12) as:

$$f_2(p, g(S)) \times g'(S)F = C_1(S, 1). \tag{S-1.13}$$

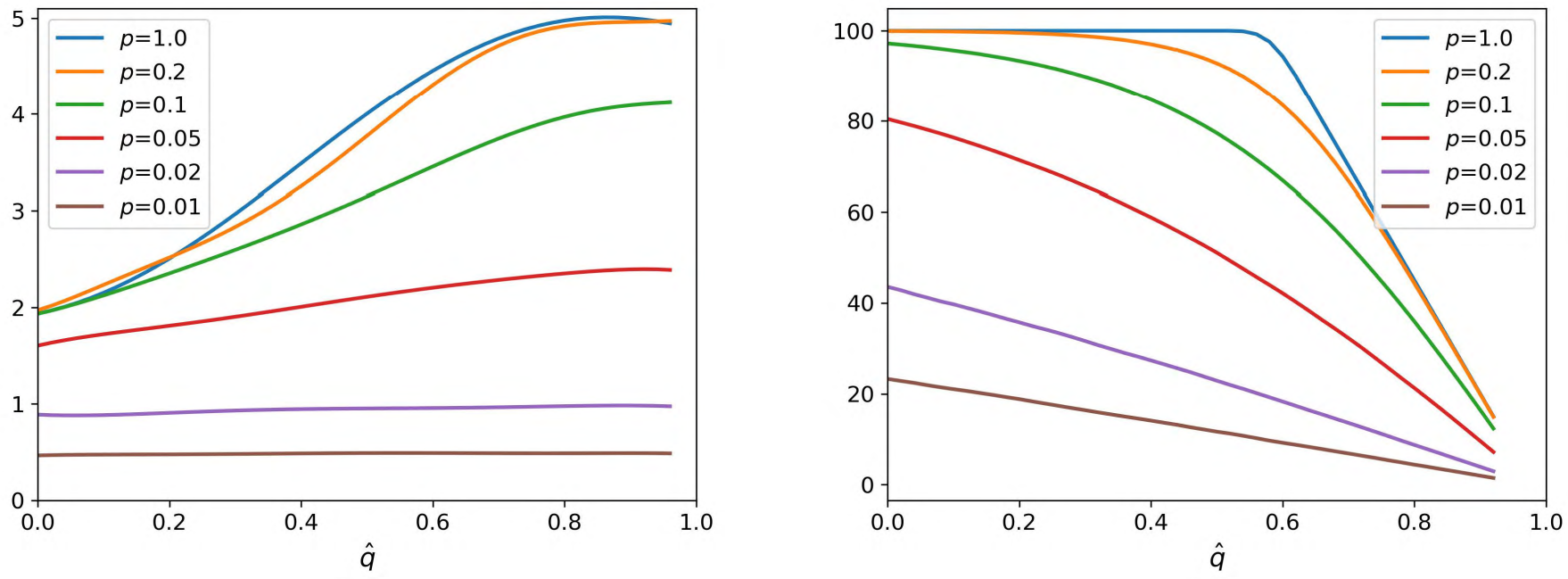

(a) (b)

Fig. S2: Illustration of imperfect social matching for a single systematic round of encounters. (a): Number $f(p, \hat{q})$ of followers a signaler might expect, depending on the signaling threshold $\hat{q}$. (b) Total number of non-random following links depending on $\hat{q}$. Parameters are $N = 50$, $k = 5$, $m = 2$.

Equation (S-1.13) has a solution only if $f_2(p, \hat{q})$ is non-zero. As we can infer from Figure S2 (a), this requires $p$ to be large enough ($p \gtrsim 0.05$ for the parameters of the figure), otherwise we end up in a non-signaling situation.

# S-2 Second-order signals

## S-2.1 Praising is an ESS

We suppose that agents may send a second-order signal by praising individuals who sent the most intense signal among those they encountered. The point of praising is to draw attention to one's own signal, shifting its visibility from $p_1$ to $p_2$. We note $\hat{q}_1$ and $\hat{q}_2$ the signaling thresholds without and with praising. Signalers will be tempted to praise as long as the cost of praising $c_p$ is small enough:

$$(f(p_2, \hat{q}_1) - f(p_1, \hat{q}_1)) \times F > c_p. \tag{S-2.1}$$

Praise is viable for all signalers if:

$$(f(p_2, \hat{q}_2) - f(p_1, \hat{q}_2)) \times F > c_p. \tag{S-2.2}$$

When this condition is met, all signalers praise to avoid being mistaken for non-signalers, while non-signalers spare the cost of praising by not praising. This all-praise situation among signalers is evolutionarily stable. If we suppose that $p_2 > m/k$, which means that praising is efficient as a second-order signal, then $f(p_2, \hat{q}_2) = k$. The overall effect of general praising is to shift the signaling threshold from $\hat{q}_1$ up to $\hat{q}_2 = 1 - \frac{m}{k}$. Writing that payoff is zero close to the threshold, we get a condition from which the new value $S_2$ of the signal can be retrieved:

$$C(S_2, 1 - \frac{m}{k}) + c_p = kF. \tag{S-2.3}$$

Note also that praising not only becomes an obligatory behavior, but represents an additional handicap for signalers as they get no relative advantage from it.

## S-2.2 Emergence of heroes

Praising has a side-effect which is to encourage a few individuals to compete for being praised and receive a bonus $s$ for each admirer they impress. Contrary to the affiliation game, the number of admirers is only limited by the size $N$ of the group. The competition leads the most competent individuals to send disproportionate signals. We note $P = N(1 - \hat{q}_2)s$ the total available payoff due to praise (if $p_2 > m/k$, then $P = Nms/k$). Under certain conditions, one may observe a winner-take-all situation in which the most competent individual is the target of all praising acts. We explore the conditions that allow $H > 1$ heroes (i.e. praised individuals) to emerge.

A multi-hero situation occurs if the most competent individual has no incentive to increase its signal $S_H$. This occurs if the minimal noticeable relative increase $\delta$ involves a prohibitive cost:

$$(P - \frac{P}{H}) < C(S_H(1+\delta), 1) - C(S_H, 1). \quad \text{(S-2.4)}$$

A first-order approximation of this condition when $\delta$ is small reads:

$$P(1 - \frac{1}{H}) < \delta S_H \times C_1(S_H, 1). \quad \text{(S-2.5)}$$

(S-2.5) can only be satisfied if $C_1(S_H, 1)$ reaches large values. This may occur if signaling cost increases asymptotically as the signal approaches a limit $L$. We will make the following hypotheses to illustrate the phenomenon:

- signaling cost is inversely proportional to competence $q$;
- individuals measure the relative distance to $L$ when evaluating the cost of increasing the signal, i.e. the marginal cost following a tiny variation $dS$ of the signal is proportional to $\frac{dS}{L-S}$.

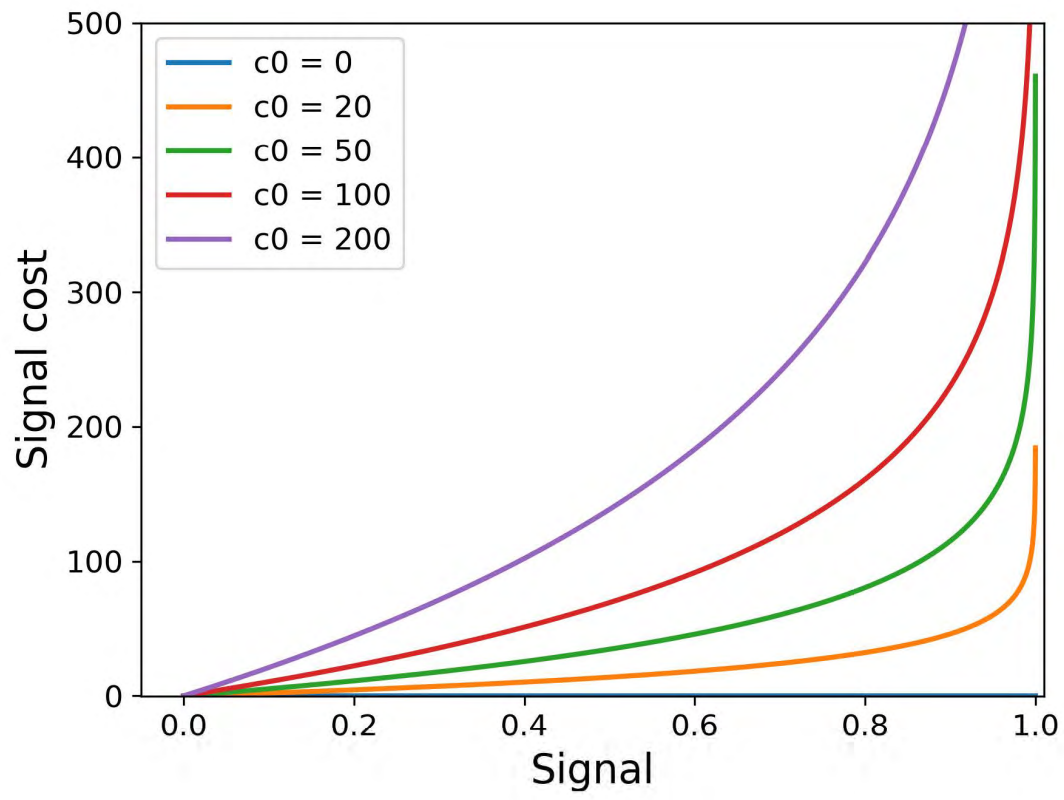

Fig. S3: Logarithmic cost $C(S, 1)$ when signals are bounded: $S < 1$.

With these assumptions, the first partial derivative of the signaling cost close to the limit $L$ has the form:

$$C_1(S, q) = \frac{c_0}{q(L - S)}, \quad \text{(S-2.6)}$$

where the cost coefficient $c_0$ controls the concavity of the cost curve. The cost turns out to be logarithmic in $S$:

$$C(S, q) = \frac{c_0}{q} \log \frac{1}{1 - \frac{S}{L}}, \quad \text{(S-2.7)}$$

Figure S3 illustrates (S-2.7) for $L = 1$ and $q = 1$. Using (S-2.6), condition (S-2.5) becomes:

$$S_H \geq \frac{L}{1 + \frac{c_0 \delta}{P(1-\frac{1}{H})}}. \quad \text{(S-2.8)}$$

The "super-signal" sent by heroes should be affordable to the least competent/motivated among them. With our assumption that competence is uniformly distributed, the competence of this $H^{th}$ hero is $1 - \frac{H}{N}$. The condition for affordability reads:

$$
\begin{aligned}
kF + \frac{P}{H} &\geq C(S_H, 1 - \frac{H}{N}) \\
&= \frac{c_0}{1 - \frac{H}{N}} \log \frac{1}{1 - \frac{1}{1 + \frac{c_0 \delta}{P(1 - \frac{1}{H})}}} \\
&= \frac{c_0}{1 - \frac{H}{N}} \log(1 + \frac{P}{c_0 \delta}(1 - \frac{1}{H})).
\end{aligned}
\tag{S-2.9}
$$

Figure 3 in the main article shows how the value of $H$ (when larger than 1) varies as a function of $c_0$ and $N$.

# S-3 Implementation

The multi-agent simulation, written in Python, is based on the *Evolife* platform[1]. Agents differ by their competence, which is uniformly distributed between 0 and 100. They learn a feature named *PraiseProbability* and decide to engage in praising accordingly. They invest in a signal intensity that depends on another feature named *Signal*; this entails a cost that smoothly decreases with their competence (function $D$) and increases asymptotically when the signal gets close to 100 (algorithm 0). They learn these two features through a simple local search. A typical example of run can be seen on the website.

**Algorithm 0** Initialization

**Input:** $self$
$self.Praising \leftarrow (random() \leq self.PraiseProbability)$
$self.SignalCost \leftarrow D(self.competence) \times -\alpha \log(1 - self.Signal/100)$
▷▷ *(D is a drecreasing function of competence)*

**Algorithm 1** Observe

```
Input: self, Partner
if random() ≤ InitialVisibility then
    if self.signal < Partner.signal then
        ▷▷ Partner remembered as potential praise target
        add (Partner, Partner.signal) to self's praise memory
    end if
    ▷▷ Partner remembered as potential affiliation target
    add (Partner, Partner.signal) to self's affiliation memory
end if
```

All interactions in the simulation are meant to be local. Individuals meet each other systematically in a randomized order. During their first encounter (algorithm 1), they observe each other's signal with a certain probability which depends on a global parameter called *InitialVisibility* (parameter $p_1$ in the model).

During a second randomized encounter (algorithm 2), individuals may praise third parties. This behavior may lead the partner to see the praiser's signal with a given probability *PraiseVisibility* (parameter $p_p$ in the model).

In a third randomized encounter, individuals select a given number ($m$ in the model) of signalers to affiliate with (algorithm 3).

After these three rounds, payoffs are computed (algorithm 4): individuals get rewarded for having attracted affiliates (they receive *FollowerImpact*, which corresponds to parameter $F$ in the model). Individuals get a bonus each time they were the target of praise (parameter $s$ in the model). Then they pay the cost of signaling. *SignalCostCoef* corresponds to $c_0$ and *PraiseCostCoef* to $c_p$ (note that praising cost is attached to *PraiseProbability* and not to actual praising). Agents' memory is reset after the assessment phase. However, agents store payoffs and learn periodically from them. They have a limited lifespan and get fully reinitialized when being reborn with the same competence.

[1] All programs are open source and are available at this Website: https://evolife.telecom-paris.fr/praise.

**Algorithm 2** Praise

```
Input: self, Partner
if self.Praising then
    ▷▷ self communicates about best encountered individual
    Target ← best signaling agent in self's praise memory
    if Target.signal > Partner.signal then
        add (Target, Target.signal) to Partner's praise memory
    end if
    ▷▷ self becomes more visibile to Partner
    if random() ≤ PraiseVisibility then
        add (self, self.signal) to Partner's affiliation memory
    end if
end if
```

**Algorithm 3** Interact

```
Input: self, Partner
if  Partner in self's affiliation memory
or  Partner in self's praise memory then
    PartnerSignal ← Partner's memorized signal
else
    PartnerSignal ← 0
end if
if  self's affiliation list is not full
or  PartnerSignal ≥ self's worst previous partner's signal then
    ▷▷ self affiliates to Partner
    ▷▷ as a result, worst previous partner is forgotten (if necessary)
    self.affiliate(Partner, PartnerSignal)
end if
```

**Algorithm 4** Assessment

```
Input: self
for A in self's affiliates do
    ▷▷ payoff for having attracted a follower
    self.Points +← FollowerImpact
end for
▷▷ paying costs
self.Points −← SignalCostCoef × self.SignalCost
self.Points −← PraiseCostCoef × σ(self.PraiseProbability)
▷▷ (σ is a sigmoid function)
if self.Praising then
    Target ← best signaling individual in self's praise memory
    ▷▷ praise target is rewarded
    Target.Points +← PraiseBonus
end if
self.resetMemory()
```

| *Description* | *Variable* | *Default value* |
|---|---|---|
| Population size | $N$ | 50 |
| Number of affiliations (followees) | $m$ | 2 |
| Number of followers | $k$ | 5 |
| Impact of being followed | $F$ | 10 |
| Impact of praise | $s$ | 3 |
| Signaling cost coefficient | $c_0$ | 100 |
| Signaling cost decrease | | 3 |
| Praise cost | $c_p$ | 3 |
| Signal precision | $\delta$ | 0.05 |
| Initial visibility | $p_1$ | 0.1 |
| Visibility due to praise | $p_p$ | 0.3 |

Table S1: List of most relevant parameters.

Due to locality and chance, there is a variance in the number of affiliates each visible signaler may attract (see section S-1.3). Another source of variance comes from the fact that agents do not always adopt the ideal strategy corresponding to their competence. They need time to learn their various options (sending the signal, expressing praise) and they constantly explore alternatives with a certain probability. Despite behavioral variance due to chance and to this "learning noise", the simulation is robust, i.e. it produces similar outcomes for a wide range of parameter values.

## S-3.1 Parameters

The simulation program relies on several parameters. The most relevant ones are listed in table S1. 'Population size' corresponds to the variable $N$ in the model. Individuals are allowed to affiliate with up to 'Number of affiliations' ($m$) other individuals, and receive up to 'Number of followers' ($k$) affiliations. Note that it is essential that $k > m$ to create a social competition. Individuals receive 'Impact of being followed' ($F$) for each of their follower. 'Impact of praise' ($s$ in the model) is given to individuals each time they are someone's praise target. 'Signaling cost coefficient' provides the scale of signal cost: it corresponds to the the cost $c_0$ paid by a medium-quality signaler. 'Signaling cost decrease' controls the variation of signaling cost depending on competence (function $D$ in algorithme 0) (0: no variation; 1: linear decrease; higher values: steeper, non-linear convex decrease). The parameter 'Praise cost' implements a gradual version of model's fixed cost $c_p$: praising individuals pay a cost that is roughly proportional to the (learned) propensity to express praise (see alg. 4). Algorithms 1, 2 and 3 involve signal comparisons. Signals are rounded to 'Signal precision' ($\delta$) to ensure meaningful comparisons. 'Initial visibility' is the probability of individuals' signals to be seen during the observation round ($p_1$ in the model).

Figure S4 shows an example of simuation run. We can observe in (a) the segmentation of the population into three types of signalers. Normal signalers in (b) are scattered over a few signal levels due to the learning noise inherent to this type of simulation. Praising probability among signalers rises rapidly toward its target value of 0.75 (orange curve).

Figures S5 and S6 explore the role of some of these parameters, while non-varying parameters are set to the default values of table S1. Figure S5 shows that the presence of praise tends to promote high-cost signaling. Figure S6 (a) shows that top signalers compete more when social stakes $k \times F$ are high. In (b), we verify that praise is maximal when the competition for attracting followers is harsh ($k$ small). Note that when $k \leq m$ (here $m = 2$), there is no social competition whatsoever. Parameters 'Signal cost decrease' and 'Signal precision' have no significant effects when kept within reasonable range.

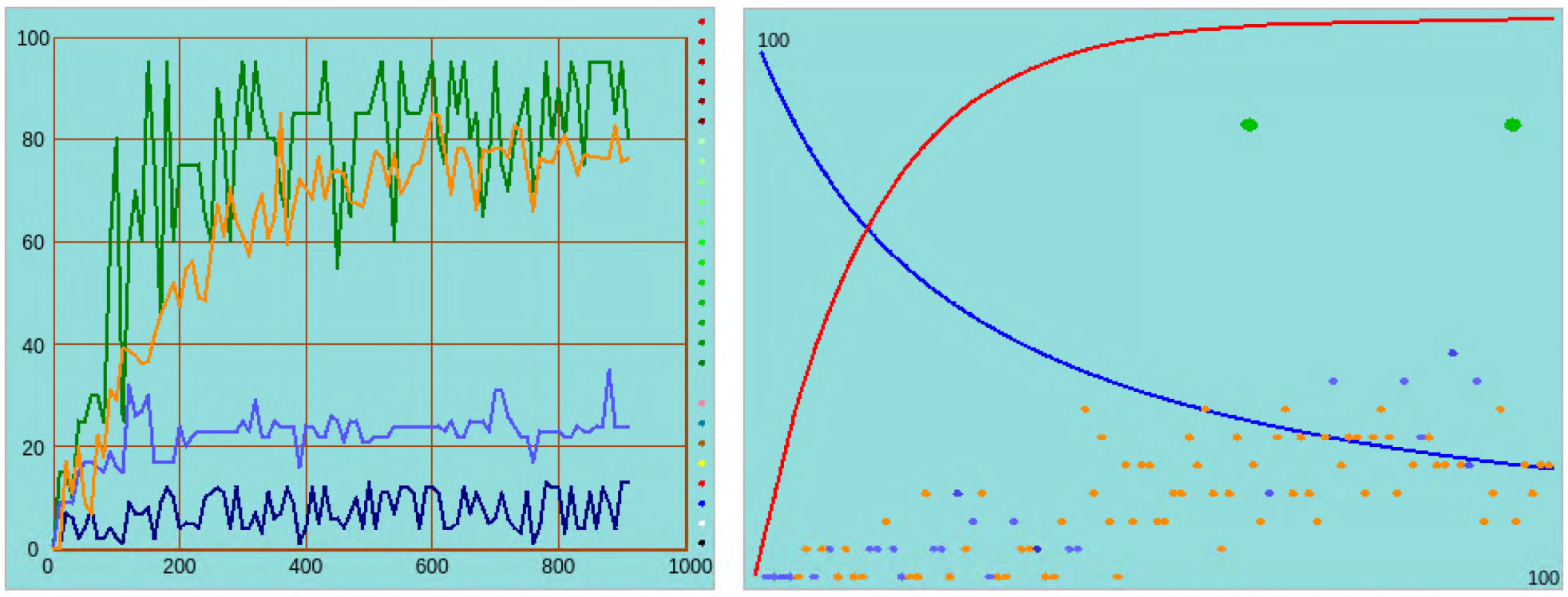

(a) Signal levels and praise probability (b) Signal levels and cost functions

Fig. S4: Example of simulation run. Parameters values are those of Table S1, except for $N = 100$.
(a) Evolution of signals for the lower (dark blue) and the higher (blue) signaling groups, and for top signalers (green); praise probability of signalers (orange).
(b) Signals depending on competence at time step 920. Green dots are praised individuals; orange dots are praising individuals. The blue decreasing curve is the signal cost coefficient depending on competence (function $D = C(., q)$) divided by 5 (actual cost for $q = 50$ is 1.65). The red increasing curve indicates how the signal cost coefficient (x-axis) grows asymptotically depending on signal level (y-axis) (function $C(S, .)$).

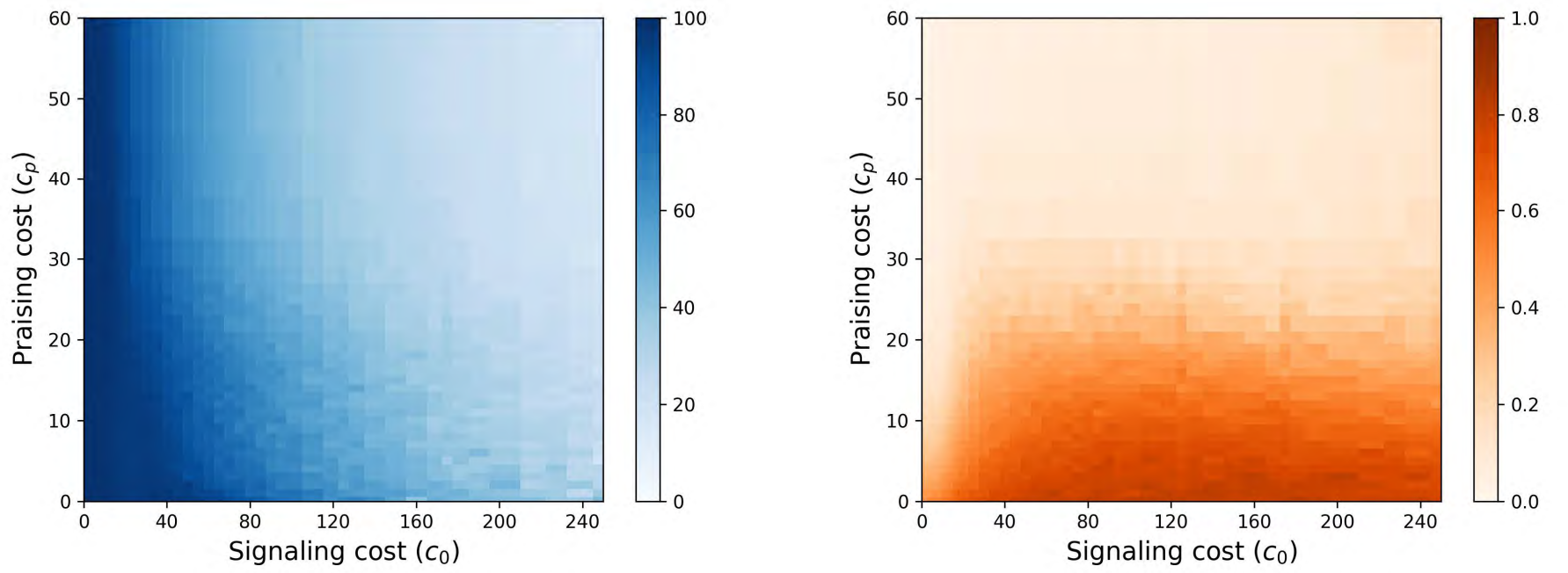

(a) Top signal intensity (b) Praise probability

Fig. S5: Impact of costs.
Signal displayed by top signalers (heroes) (a) and praise probability (b) when costs vary. Parameters values are those of Table S1.

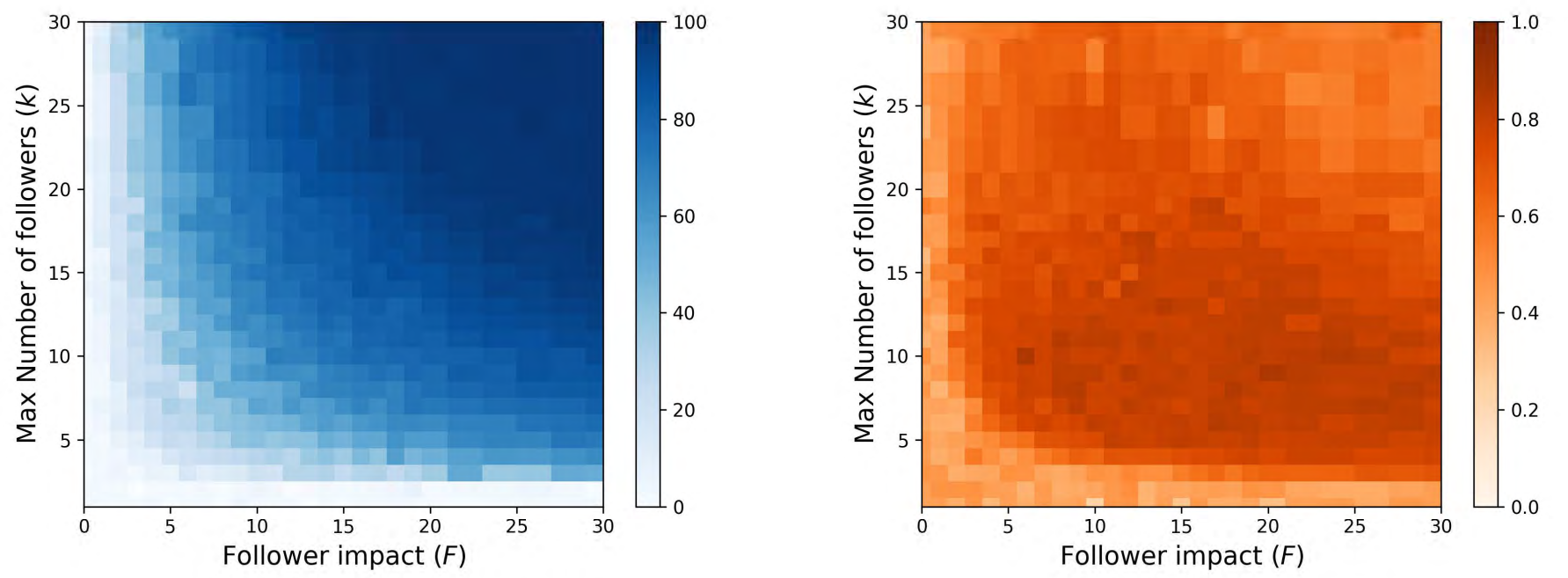

(a) Top signal intensity (b) Praise probability

Fig. S6: Impact of social stakes.
Signal displayed by top signalers (heroes) (a) and praise probability (b) depending on 'Follower impact' ($F$) and 'Number of followers' ($k$). Parameters values are those of Table S1, except for $c_0 = 200$.